\documentclass[12pt,a4paper]{article}
\usepackage{amsfonts}
\usepackage{amssymb}

\def\hybrid{\topmargin -20pt    \oddsidemargin 0pt
        \headheight 0pt \headsep 0pt
        \textwidth 6.25in       
        \textheight 9.5in       
        \marginparwidth .875in
        \parskip 5pt plus 1pt   \jot = 1.5ex}

\hybrid

\def\baselinestretch{1.2}

\catcode`\@=11

\def\marginnote#1{}
\newcount\hour
\newcount\minute
\newtoks\amorpm
\hour=\time\divide\hour by60
\minute=\time{\multiply\hour by60 \global\advance\minute by-\hour}
\edef\standardtime{{\ifnum\hour<12 \global\amorpm={am}%
        \else\global\amorpm={pm}\advance\hour by-12 \fi
        \ifnum\hour=0 \hour=12 \fi
        \number\hour:\ifnum\minute<10 0\fi\number\minute\the\amorpm}}
\edef\militarytime{\number\hour:\ifnum\minute<10 0\fi\number\minute}

\def\draftlabel#1{{\@bsphack\if@filesw {\let\thepage\relax
   \xdef\@gtempa{\write\@auxout{\string
      \newlabel{#1}{{\@currentlabel}{\thepage}}}}}\@gtempa
   \if@nobreak \ifvmode\nobreak\fi\fi\fi\@esphack}
        \gdef\@eqnlabel{#1}}
\def\@eqnlabel{}
\def\@vacuum{}
\def\draftmarginnote#1{\marginpar{\raggedright\scriptsize\tt#1}}

\def\draft{\oddsidemargin -.5truein
        \def\@oddfoot{\sl preliminary draft \hfil
        \rm\thepage\hfil\sl\today\quad\militarytime}
        \let\@evenfoot\@oddfoot \overfullrule 3pt
        \let\label=\draftlabel
        \let\marginnote=\draftmarginnote
   \def\@eqnnum{(\theequation)\rlap{\kern\marginparsep\tt\@eqnlabel}%
\global\let\@eqnlabel\@vacuum}  }

\def\preprint{\twocolumn\sloppy\flushbottom\parindent 2em
        \leftmargini 2em\leftmarginv .5em\leftmarginvi .5em
        \oddsidemargin -.5in    \evensidemargin -.5in
        \columnsep .4in \footheight 0pt
        \textwidth 10.in        \topmargin  -.4in
        \headheight 12pt \topskip .4in
        \textheight 6.9in \footskip 0pt
        \def\@oddhead{\thepage\hfil\addtocounter{page}{1}\thepage}
        \let\@evenhead\@oddhead \def\@oddfoot{} \def\@evenfoot{} }

\def\numberbysection{\@addtoreset{equation}{section}
        \def\theequation{\thesection.\arabic{equation}}}

\def\underline#1{\relax\ifmmode\@@underline#1\else
        $\@@underline{\hbox{#1}}$\relax\fi}

\def\titlepage{\@restonecolfalse\if@twocolumn\@restonecoltrue\onecolumn
     \else \newpage \fi \thispagestyle{empty}\c@page\z@
        \def\thefootnote{\fnsymbol{footnote}} }

\def\endtitlepage{\if@restonecol\twocolumn \else \newpage \fi
        \def\thefootnote{\arabic{footnote}}
        \setcounter{footnote}{0}}  

\catcode`@=12
\relax

\def\figcap{\section*{Figure Captions\markboth
        {FIGURECAPTIONS}{FIGURECAPTIONS}}\list
        {Figure \arabic{enumi}:\hfill}{\settowidth\labelwidth{Figure
999:}
        \leftmargin\labelwidth
        \advance\leftmargin\labelsep\usecounter{enumi}}}
 \relax
\def\tablecap{\section*{Table Captions\markboth
        {TABLECAPTIONS}{TABLECAPTIONS}}\list
        {Table \arabic{enumi}:\hfill}{\settowidth\labelwidth{Table
999:}
        \leftmargin\labelwidth
        \advance\leftmargin\labelsep\usecounter{enumi}}}
 \relax
\def\reflist{\section*{References\markboth
        {REFLIST}{REFLIST}}\list
        {[\arabic{enumi}]\hfill}{\settowidth\labelwidth{[999]}
        \leftmargin\labelwidth
        \advance\leftmargin\labelsep\usecounter{enumi}}}
 \relax
\makeatletter
\newcounter{pubctr}
\def\publist{\@ifnextchar[{\@publist}{\@@publist}}
\def\@publist[#1]{\list
        {[\arabic{pubctr}]\hfill}{\settowidth\labelwidth{[999]}
        \leftmargin\labelwidth
        \advance\leftmargin\labelsep
        \@nmbrlisttrue\def\@listctr{pubctr}
        \setcounter{pubctr}{#1}\addtocounter{pubctr}{-1}}}
\def\@@publist{\list
        {[\arabic{pubctr}]\hfill}{\settowidth\labelwidth{[999]}
        \leftmargin\labelwidth
        \advance\leftmargin\labelsep
        \@nmbrlisttrue\def\@listctr{pubctr}}}
 \relax
\makeatother
\newskip\humongous \humongous=0pt plus 1000pt minus 1000pt

\newif\ifdtup

\relax

\def\be{\begin{equation}}
\def\ee{\end{equation}}
\def\ba{\begin{eqnarray}}
\def\ea{\end{eqnarray}}

\def\om{\omega}

\def\no{\noindent}

\def\qq{\qquad}

\def\IR{\relax{\rm I\kern-.18em R}}

\def\IR{\relax{\rm I\kern-.18em R}}
\def\inv{^{\raise.15ex\hbox{${\scriptscriptstyle -}$}\kern-.05em 1}}

\def\tL{{\tilde L}}

\begin{document}

\renewcommand{\theequation}{\arabic{equation}}

\newcommand{\beq}{\begin{equation}}
\newcommand{\eeq}[1]{\label{#1}\end{equation}}
\newcommand{\ber}{\begin{eqnarray}}
\newcommand{\eer}[1]{\label{#1}\end{eqnarray}}
\newcommand{\eqn}[1]{(\ref{#1})}
\begin{titlepage}
\begin{center}

\hfill CERN-TH/2000-046\\
\vskip -.1 cm
\hfill NEIP-00-003\\
\vskip -.1 cm
\hfill February 2000\\
\vskip -.1 cm
\hfill hep--th/0002092\\

\vskip .2in

{\large \bf Riemann surfaces and Schr\"odinger potentials\\ 
of gauged supergravity}\footnote{Contribution to the proceedings of
the TMR meeting {\em Quantum Aspects of Gauge Theories, Supersymmetry
and Unification} held in Paris, September 1999 (to appear in the
JHEP proceedings section); based on two talks presented by I.B. and
K.S.}

\vskip 0.3in

{\bf I. Bakas${}^1$},\phantom{x} {\bf A. Brandhuber${}^2$} \phantom{x} 
and \phantom{x} {\bf K. Sfetsos${}^3$}
\vskip 0.1in
{\em ${}^1\!$Department of Physics, University of Patras\\
GR-26500 Patras, Greece\\
\footnotesize{\tt bakas@nxth04.cern.ch, ajax.physics.upatras.gr}}\\
\vskip0.1in
{\em ${}^2\!$Theory Division, CERN\\
CH-1211 Geneva 23, Switzerland\\
\footnotesize{\tt brandhu@mail.cern.ch}}\\
\vskip0.1in
{\em ${}^3\!$Institut de Physique, Universit\'e de Neuch\^atel\\
Breguet 1, CH-2000 Neuch\^atel, Switzerland\\
\footnotesize{\tt sfetsos@mail.cern.ch}}\\

\end{center}

\vskip .3in

\centerline{\bf Abstract}

\no
Supersymmetric domain-wall solutions of maximal gauged supergravity
are classified in 4, 5 and 7 dimensions in the presence of non-trivial
scalar fields taking values in the coset $SL(N, \IR)/SO(N)$ for 
$N=8,~6$ and 5 respectively. We use an algebro-geometric method based on
the Christoffel--Schwarz transformation, which allows for the characterization
of the solutions in terms of Riemann surfaces whose genus depends on the
isometry group. The uniformization of the curves can be carried out 
explicitly for models of low genus and results into trigonometric and
elliptic solutions for the scalar fields and the conformal factor of the 
metric. The Schr\"odinger potentials for the quantum fluctuations of the
graviton and scalar fields are derived on these backgrounds and enjoy
all properties of supersymmetric quantum mechanics. Special attention is 
given to a class of elliptic models whose quantum fluctuations are  
commonly described by the generalized Lam\'e potential
$\mu (\mu +1){\cal P}(z) + \nu (\nu +1){\cal P}(z+\omega_1) + 
\kappa (\kappa +1){\cal P}(z+\omega_2) + 
\lambda (\lambda +1){\cal P}(z+\omega_1 +\omega_2)$ 
for the Weierstrass
function ${\cal P}(z)$ of the underlying Riemann surfaces with periods 
$2\omega_1$ and $2\omega_2$, and for different half-integer values
of the coupling constants $\mu,\nu,\kappa,\lambda$.

\vfill
\end{titlepage}
\eject

\def\baselinestretch{1.2}
\baselineskip 16 pt
\noindent

\def\tT{{\tilde T}}
\def\tg{{\tilde g}}
\def\tL{{\tilde L}}


\noindent
Recently there has been some progress towards the construction
of supersymmetric domain-wall solutions of $D$-dimensional
gravity coupled to scalar fields taking values in the coset
space $SL(N, \IR)/SO(N)$. There are three cases of particular
interest in the context of maximally gauged supergravities, namely
$(D,~N)=(4, ~8), ~(5, ~6)$ and $(7, ~5)$. In this contribution
we summarize the results obtained in the subject by reducing the
classification and construction of all such domain-wall 
solutions to a problem of algebraic geometry, where Riemann 
surfaces arise naturally in connection with the 
Christoffel--Schwarz transformation in complex analysis.
In the generic case,
where no continuous subgroup of the original $SO(N)$ gauge symmetry 
remains unbroken, we find that the algebraic curve of the corresponding
solution is a Riemann surface of genus $N+1$ depending on $N$ real
moduli. When some cycles shrink to zero size, by letting some of the
moduli coalesce, the symmetry group is enhanced, whereas the genus
of the underlying Riemann surface is lowered accordingly. It is then 
appropriate to think of the breaking of $SO(N)$ to its various 
subgroups, 
which remain and characterize the individual solutions, as in 
the case of spontaneous symmetry breaking. The explicit
construction of the domain-walls amounts to the uniformization of 
the algebraic curves, which can be easily carried out in the cases
of low genus (0 or 1); as a result, the simplest solutions can be  
described in terms of elementary functions (rational or trigonometric)
for genus 0 surfaces 
or in terms of elliptic functions for genus 1 surfaces.    

A preliminary version of these results was announced last September
in the TMR meeting on ``Quantum Aspects of Gauge Theories, 
Supersymmetry and Unification" held in Paris,  
but a more extensive account was subsequently given in two recent papers 
\cite{bs1, bs2}, where the reader can find more details and a list of 
references to other work on this interesting topic of
research. There are alternative ways to describe our solutions by 
lifting them to the string/M-theory frame, thus showing that they are
consistent truncations of ten- or eleven-dimensional supergravity. 
In particular, from an eleven-dimensional point of view, the 
domain-wall solutions of $D=4$ and $D=7$ gauged supergravities correspond
to various continuous distributions of M2- and M5-branes respectively.
Likewise, from the point of view of ten-dimensional type-IIB 
supergravity, the domain-wall solutions of $D=5$ gauged supergravity  
correspond to the gravitational field of D3-branes continuously 
distributed on hyper-surfaces embedded in the six-dimensional space
transverse to the branes. Such higher dimensional backgrounds arise
in the context of the AdS/CFT correspondence as supergravity duals of 
the field theories living on M2, M5 or D3 branes on the Coulomb branch,
and therefore they have numerous applications. Domain-wall solutions
of gauged supergravities are also interesting to consider in relation
to proposals that view our world as a membrane embedded 
non-trivially in a higher dimensional space-time, and for which there is
a normalizable graviton zero mode, 
like in the recent scenario of Randall and Sundrum where a resolution of the 
mass hierarchy problem in geometrical terms was proposed.

In the present work we will focus mostly on the systematic description
of the scalar and graviton field fluctuations on the domain-wall 
backgrounds, apart from reviewing the general 
algebro-geometric aspects of the corresponding solutions in gauged 
supergravities for $D=4, ~5$ and 7. It turns out that the spectrum of
these quantum fluctuations can be formulated as a simple problem 
in one-dimensional supersymmetric quantum mechanics, where the 
Schr\"odinger potential is written in terms of a prepotential given
(up to a scale) 
by the conformal factor of the underlying domain-wall metric. However,
the exact form of the spectrum is very difficult to obtain beyond the
WKB approximation, in particular for those solutions that correspond
to genus 1 (or higher) Riemann surfaces. 
Nevertheless, in many cases corresponding to genus 0 Riemann surfaces, the 
exact spectrum can be computed explicitly.
We will discuss some  
aspects of the Schr\"odinger potentials that arise in several elliptic
models, and which have the  form of generalized Lam\'e potentials 
with half-integer characteristics. Thus, 
finite-zone potentials, which are familiar from the study of the
KdV equation and the hierarchies of non-linear differential equations
descending from it, do not arise in theories of gauged supergravity
and as a result there is only very little known now about the 
exact spectrum.
Of course, one might construct domain-wall solutions of $D$-dimensional
gravity coupled to a selection of scalar fields outside the scope of  
gauged supergravities, which could yield finite-zone potentials, but  
these models would not arise as
consistent truncations of ten- or eleven-dimensional
supergravity. These issues pose many interesting questions that deserve
further study. 
       
We consider the bosonic sector of gauged supergravity in $D$-dimensions, 
which contains only scalar fields in the coset $SL(N, \IR)/SO(N)$, for
the specific values of $D$ and $N$ that arise in consistent truncations 
of  ten- or eleven-dimensional supergravity. Since
all other fields are set to zero, the Lagrangian assumes the 
form  
\be
{\cal L} = {1 \over 4} {\cal R} - {1 \over 2} \sum_{I=1}^{N-1} 
(\partial \alpha_I)^2 - P(\alpha_I) ~, 
\ee
where the potential $P(\alpha_I)$ has the special form
\be
P(\alpha_I) = {g^2 \over 8} \left(\sum_{I=1}^{N-1} \left(
{\partial W \over \partial \alpha_I} \right)^2 - 2 {D-1 \over D-2} 
W^2 \right) ~.  
\ee
The function $W$ is more easily described in terms of $N$ real scalar
fields $\beta_i$, which are constrained to satisfy the relation
\be 
\beta_1 + \beta_2 + \cdots + \beta_N = 0 ~,\label{constr} 
\ee
as follows  
\be
W = -{1 \over 4} \sum_{i=1}^{N} e^{2 \beta_i} ~. 
\ee
The precise relation among the fields $\alpha_I$ and $\beta_i$ is 
given by
\be
\beta_i = \sum_{I=1}^{N-1} \lambda_{iI} \alpha_I ~, 
\ee
where $\lambda_{iI}$ are the elements of an $N \times (N-1)$ matrix. 
Its rows correspond to the $N$ weights 
of the fundamental representation of $SL(N)$, and as such they satisfy
the normalization conditions
\be
\sum_{I=1}^{N-1} \lambda_{iI} \lambda_{jI} = 2 \delta_{ij} - 
{2 \over N} ~, ~~~~ \sum_{i=1}^{N} \lambda_{iI} \lambda_{iJ} = 
2 \delta_{IJ} ~, ~~~~ \sum_{i=1}^{N} \lambda_{iI} = 0 ~.  
\ee
The coefficient $g^2$ appearing in front of the potential $P(\alpha_I)$ 
defines an associated length scale $R$ given by the relation 
$g=2/R$. Having said this we will set $g=1$ in the following.  

The domain-wall solutions exhibit $(D-1)$-dimensional Poincar\'e 
invariance, namely
\ba
ds^2 & = & e^{2A(z)}(\eta_{\mu \nu} dx^{\mu} dx^{\nu} + dz^2) \nonumber\\
     & = & e^{2A(r)} \eta_{\mu \nu} dx^{\mu} dx^{\nu} + dr^2 ~, 
\ea
where the metric is given in terms of a single function $A$ which depends 
on the variable $z$ or $r$, as denoted above; 
these coordinates are related by the differential 
relation $dr = - e^A dz$. Furthermore, we assume that the scalars 
depend only on $z$ (or $r$) as well. 
A careful analysis of the problem shows that
supersymmetric solutions of this kind satisfy a system of first-order
differential equations 
\be
{dA \over dr} = -{1 \over D-2} W ~, ~~~~~ 
{d \alpha_I \over dr} = {1 \over 2} {\partial W \over \partial \alpha_I} ~.
\ee
These can be obtained either directly from the Killing spinor equations or
as a saddle point of the action functional by the method of Bogomol'nyi.
In any case, it is convenient to work with the unconstrained fields
$\beta_i$ as functions of $z$, in terms of which the first-order 
equations become
\be
A^{\prime} = {1 \over D-2} e^A W ~, ~~~~ 
\beta_{i}^{\prime} = 2 {D-2 \over N} A^{\prime} + {1 \over 2} 
e^{2 \beta_i + A} ~; ~~~~ i = 1, 2, \ldots , N ~,
\ee
where the prime denotes derivative with respect to $z$.  

It is fairly easy to integrate the differential equations for the 
scalar fields $\beta_i$ by introducing an auxiliary function $F(z)$,  
which is related to the conformal factor as  
\be
e^{A(z)} = (-F^{\prime})^{{N \over 4(D-2) +N}} ~. 
\ee
Then, by simple integration, 
we obtain the following result for the scalar fields, 
\be
e^{2 \beta_i (z)} = {(-F^{\prime})^{\Delta / N} \over F - b_i} ~;  
~~~~ i = 1, 2, \cdots , N ~,  
\ee
where $\Delta = 4(D-2)N /(4(D-2) + N)$ and $b_i$ are integration
constants. Actually, we have $\Delta = 4$ for all three cases of interest
in gauged supergravity, namely $(D, N) = (4, 8)$, $(5, 6)$ and 
$(7, 5)$. Moreover, reality of the scalar fields $\beta_i$ requires
that for real $z$ we have $F(z) \geq b_{\rm max}$, where $b_{\rm max}$
is the maximum value of the real moduli $b_i$; they may be ordered 
as $b_1 \geq b_2 \geq \cdots \geq b_N$ without loss of generality. 
Taking now into account the algebraic constraint \eqn{constr} 
imposed on the 
scalar fields $\beta_i$ we arrive at the following non-linear 
differential equation for the unknown function $F(z)$ 
\be
\left(F^{\prime}(z)\right)^4 = \prod_{i=1}^{N} (F(z) - b_i) ~,
\label{master}  
\ee
that captures the non-linear aspects of the corresponding equations 
for the domain-walls for $N=8, ~6$ or 5. In fact, we will look for
solutions which are asymptotic to $AdS_D$ space with radius 
$4(D-2)/N$ (in units where $g=2/R=1$) 
as $F \rightarrow + \infty$ (or equivalently 
$z \rightarrow 0^+$), when the moduli $b_i$ assume
arbitrary values. In the special case that all moduli are equal, 
the corresponding solution is simply $AdS_D$ space
(not only asymptotically) with all real scalar fields 
being zero. Hence, classifying the solutions of equation 
\eqn{master}, 
which depend on $N$ real moduli, will provide us with the list of 
all supersymmetric domain-wall solutions in question. 
As it turns out, this problem
can be addressed systematically in the context of algebraic geometry.  

The underlying mathematical structure for solving the differential
equation \eqn{master},  
with arbitrary moduli $b_i$, is that of the Christoffel--Schwarz
transformation. 
It is useful to think of the variable $z$ as being 
complex, whereas $F$ taking values in the complex upper-half 
plane. Of course, appropriate restrictions have to be made at 
the end in order to ensure the reality of the variable $z$ and
hence the reality of our domain-wall solutions. 
We will treat the Christoffel--Schwarz transformation in a unified
way for all three cases of interest, namely $(D,N)=(4,8)$ (M2-branes),
$(D,N)=(5,6)$ (D3-branes), and $(D,N)=(7,5)$ (M5-branes), 
since there is a hierarchy of algebraic curves within this 
transformation that depends on the isometry groups
used for the distributions of branes in ten or eleven dimensions. 
It is useful to start with $N=8$
and consider an octagon in the complex $z$-plane, which is mapped
onto the upper-half plane via a Christoffel--Schwarz transformation
\be
{dz \over dF} = (F-b_1)^{-{\varphi}_1/\pi} 
(F-b_2)^{-{\varphi}_2/\pi} \cdots (F-b_8)^{-{\varphi}_8/\pi} ~.  
\ee
This transformation maps the vertices of the octagon 
to the points $b_1, b_2, \ldots , b_8$ on the real axis of the 
complex $F$-plane, whereas its interior 
is mapped onto the entire upper-half
$F$-plane. The variables ${\varphi}_i$ denote the exterior 
(deflection) angles of the octagon at the corresponding vertices,
which are constrained by geometry to satisfy the relation
${\varphi}_1 + {\varphi}_2 + \ldots + {\varphi}_8 = 2\pi$.
We proceed by making the canonical choice of angles
${\varphi}_1 = {\varphi}_2 = \ldots = {\varphi}_8 = \pi/4$, in 
which case we arrive at the differential equation that relates
$dz$ and $dF$: 
\be
\left({dz \over dF}\right)^4 = (F-b_1)^{-1}(F-b_2)^{-1} \cdots
(F-b_8)^{-1} ~, 
\ee
which is the equation we have to solve for the case of $D=4$ gauged
supergravity with scalar fields in the coset $SL(8, \IR)/SO(8)$. 

It is convenient at this point to introduce complex algebraic 
variables
\be
x=F(z) ~, ~~~~~ y =F^{\prime}(z) ~, 
\ee
which cast the differential equation above into the form of an 
algebraic curve 
\be
y^4 = (x-b_1)(x-b_2) \cdots (x-b_8) ~. 
\ee
This defines a Riemann surface, which is pictured geometrically 
by gluing four sheets together along their branch cuts.  
The task is to uniformize the algebraic curve by finding
another complex variable, call it $u$, so that $x = x(u)$ and 
$y = y(u)$, which resolves the problem of multi-valuedness of the 
algebraic equation above. 
Then, following the definition of $x$ and $y$ in terms of $F(z)$
and its $z$-derivative, one can apply the chain rule in order to 
obtain the function $z(u)$ by integration of 
the resulting first-order equation
\be
{dz \over du} = {1 \over y(u)}{dx(u) \over du} ~. 
\label{ssool}
\ee
Finally, by inverting the result one obtains the function $u(z)$, 
which yields $F(z)$, and hence the conformal factor of the 
corresponding domain-wall solutions, together with the solution for
the scalar fields. 
Of course, there is an integration constant
that appears in the function $z(u)$, but this can be fixed by 
requiring that the asymptotic behaviour of the domain-walls  
approach the $AdS$ geometry as $z \rightarrow 0^+$. We also note 
that there is a discrete symmetry $x \leftrightarrow -x$,
$b_i \leftrightarrow -b_i$ that leaves invariant the form of the
algebraic curve. It can be employed in order to set $F$ bigger or
equal than the maximum value of the moduli $b_i$ instead of being smaller 
or equal than the minimum value, thus insuring that $z \rightarrow 0^+$
corresponds to $F \rightarrow +\infty$ instead of $-\infty$.

The whole procedure is straightforward, but  turns out to
be cumbersome when the moduli $b_i$ take general values.
Matters simplify considerably when certain 
moduli are allowed to become equal, which effectively reduces the genus 
of the algebraic curve and leaves some of the isometries unbroken.
In general we will have models
for each continuous subgroup of the maximal isometry group $SO(8)$,
in which case the associated Riemann surface becomes
\be
y^4 = (x-b_1)^{k_1} (x-b_2)^{k_2} \cdots (x-b_k)^{k_r} ~, 
\ee
with $k_1 + k_2 + \cdots + k_r = 8$, and 
$SO(k_1) \times SO(k_2) \times \cdots \times SO(k_r)$ as isometry group.  
These surfaces will not be in an irreducible form if all the exponents
$k_i$ have a common divisor with 4. To calculate the genus, and also
proceed with their uniformization, we first bring the 
algebraic curves into the irreducible form (when this is necessary) 
\be
y^m = (x-b_1)^{a_1}(x-b_2)^{a_2} \cdots (x-b_n)^{a_n} ~, 
\ee
where the integer exponents (with $n \leq 8$) satisfy the relation 
$a_1 + a_2 + \cdots a_n = 2m$. Then, we write down the ratios
\be
{a_1 \over m} = {d_1 \over c_1} ~, ~ \cdots ~, {a_n \over m} = 
{d_n \over c_n} ~; ~~~~ {a_1 + \cdots + a_n \over m} = {d_0 \over c_0}   
\ee
in terms of relatively prime numbers and use the Riemann--Hurwitz 
relation
\be
g = 1 -m + {m \over 2} \sum_{i=0}^{n} \left(1 - {1 \over c_i} \right) 
\ee
to compute the genus $g$ (it should not be confused with the 
symbol used for the inverse
length scale $g = 2/R$, which has already been normalized to 1).

We present below the list of all 
Riemann surfaces that classify the domain-wall solutions of 
four-dimensional gauged supergravity with non-trivial scalar fields
in the coset $SL(8,\IR)/SO(8)$ by giving their genus according to
the Riemann--Hurwitz relation, their irreducible form (since in certain
cases the exponents have common factors and the curve might be 
reducible when written in its original form), as well as the corresponding 
isometry groups that determine the geometrical distribution of 
M2-branes in eleven dimensions. We have 22 models in 
total, namely:  

\begin{center}
\begin{tabular}{c|l|l}
{\hskip-3pt \em Genus} & {\em Irreducible Curve} & {\em Isometry Group}\\
\hline
 9 & $y^4=(x-b_1)(x-b_2) \cdots (x-b_7)(x-b_8)$ & None \\
\hline
 7 & $y^4=(x-b_1)(x-b_2) \cdots (x-b_6)(x-b_7)^2$ & $SO(2)$ \\
\hline
 6 & $y^4=(x-b_1)(x-b_2) \cdots (x-b_5)(x-b_6)^3$ & $SO(3)$ \\
\hline
 5 & $y^4=(x-b_1) \cdots (x-b_4)(x-b_5)^2(x-b_6)^2$ 
& $SO(2)^2$ \\
\hline
 4 & $y^4=(x-b_1)(x-b_2)(x-b_3)(x-b_4)^2(x-b_5)^3$ 
& $SO(2) \times SO(3)$ \\
\hline
 3 & $y^4=(x-b_1) \cdots (x-b_4)(x-b_5)^4$ & $SO(4)$ \\
    & $y^4=(x-b_1)(x-b_2)(x-b_3)(x-b_4)^5$ & $SO(5)$ \\ 
    & $y^4=(x-b_1)(x-b_2)(x-b_3)^3(x-b_4)^3$ 
& $SO(3)^2$ \\
    & $y^4=(x-b_1)(x-b_2)(x-b_3)^2(x-b_4)^2(x-b_5)^2$ 
& $SO(2)^3$ \\
\hline
 2 & $y^4 = (x-b_1)(x-b_2)^2(x-b_3)^2(x-b_4)^3$ 
& $SO(2)^2 \times SO(3)$ \\
\hline
 1 & $y^4=(x-b_1)(x-b_2)(x-b_3)^6$ & $SO(6)$ \\
    & $y^4= (x-b_1)(x-b_2)(x-b_3)^2(x-b_4)^4$ 
& $SO(2) \times SO(4)$ \\
    & $y^4=(x-b_1)(x-b_2)^2(x-b_3)^5$ & $SO(2) \times SO(5)$ \\
    & $y^4=(x-b_1)^2(x-b_2)^3(x-b_3)^3$ 
& $SO(2) \times SO(3)^2$ \\
    & $y^2=(x-b_1)(x-b_2)(x-b_3)(x-b_4)$ 
& $SO(2)^4$ \\
\hline
 0 & $y^4=(x-b_1)(x-b_2)^7$ & $SO(7)$ \\
    & $y = (x-b)^2$ & $SO(8)$ (maximal) \\
    & $y^2=(x-b_1)(x-b_2)^3$ & $SO(2) \times SO(6)$ \\
    & $y^4=(x-b_1)(x-b_2)^3(x-b_3)^4$ & $SO(3) \times SO(4)$ \\
    & $y^4=(x-b_1)^3(x-b_2)^5$ & $SO(3) \times SO(5)$ \\
    & $y = (x-b_1)(x-b_2)$ & $SO(4)^2$ \\
    & $y^2=(x-b_1)(x-b_2)(x-b_3)^2$ 
& $SO(2)^2 \times SO(4)$ \\
\hline
\end{tabular}
\end{center}
     
It is interesting to note that the classification of domain-walls 
of five-dimensional gauged supergravity with non-trivial scalar 
fields in the coset $SL(6, \IR)/SO(6)$ follows immediately from 
above by restricting our attention to models with 
an $SO(2)$ factor in 
the isometry group. It is clear that in this case the classification
of solutions reduces to the list of all algebraic curves  
\be
y^4 = (x-b_1)(x-b_2) \cdots (x-b_6) \ ,
\ee
depending on the values of the six real moduli $b_i$. But such curves 
can be viewed as special cases of the $N=8$ curves when
$b_7 = b_8 = - \infty$; the limiting point is taken to be $- \infty$ 
rather than $+ \infty$ in order to keep the ordering 
$b_1 \geq b_2 \geq \cdots \geq b_8$ that is usually made. 
In other words, using the geometrical framework of the 
Christoffel--Schwarz transformation, we consider that 
the octagon in the complex $z$-plane
degenerates by shrinking one of its sides to zero size (in which case 
the corresponding deflection angle becomes $\pi / 2$) and the resulting
double vertex is mapped to $-\infty$ on the real $F$-line. 
Therefore, by comparison with  
the list above, we obtain immediately the table of all domain-walls of 
five-dimensional gauged 
supergravity, which correspond to various continuous distributions
of D3-branes in ten dimensions. We have 11 such models in
total, as follows by inspection, which maintain the genus of their 
``parent" $N=8$ algebraic curves; $g$ is ranging now from 7 to 0 depending
on the isometry groups of the individual models. 
 
Finally, the algebraic classification of all domain-wall solutions
of seven-dimensional gauged supergravity with non-trivial scalar
fields in the coset $SL(5, \IR)/SO(5)$ (which by the way provides
the full scalar sector in this case) follows by considering all 
Riemann surfaces of the form
\be
y^4 = (x-b_1)(x-b_2) \cdots (x-b_5)\ ,
\ee
for various values of the five real moduli $b_i$. As before, these 
surfaces can be viewed as special cases of the $N=8$ algebraic 
curves where three of the moduli are taken to infinity, i.e. 
$b_6 = b_7 = b_8 = - \infty$, whereas the remaining are free to
vary. As before, in terms of the Christoffel--Schwarz transformation, 
the original octagon in the complex $z$-plane degenerates to an 
exagon with one of its deflection angles becoming now $3 \pi /4$, and
the resulting triple vertex is mapped to $- \infty$ on the real $F$-line. 
Put differently, we may compose the list of all domain-walls
that correspond to various continuous distributions of M5-branes
in eleven dimensions by considering all $N=8$ models with a 
$SO(3)$ isometry factor. Thus, we have 7 such models in total, which 
follow by inspection from the list above, all having the same genus as
their ``parent" $N=8$ algebraic curves; $g$ is now ranging from 
6 to 0 depending on the isometry group. 
We note for completeness in this latter case that 
the invariance of the curves under the 
discrete symmetry $x \leftrightarrow -x$, $b_i \leftrightarrow - b_i$ 
is not present any more, because the 
corresponding algebraic equations contain
only an odd number of factors.  

In summary, for generic values of the moduli parameters $b_i$, the
domain-wall solutions of $D$-dimensional gauged supergravity are
described by a Riemann surface of genus $N+1$ in the presence of 
non-trivial scalar fields in the coset space $SL(N, \IR)/SO(N)$. 
As certain cycles shrink to zero size by letting some of the 
moduli coalesce, the genus of the algebraic curve becomes smaller
and the corresponding domain-wall solutions have as symmetry the
appropriate subgroups of $SO(N)$. The explicit construction of the
solutions requires to perform the uniformization of the 
associated Riemann surfaces, which can be easily done for the cases
of low genus, namely 0 or 1. For genus 0, one has to employ 
bi-rational transformations from $x$, $y$ to new variables 
$v(x,y)$, $w(x,y)$, so that the algebraic curve assumes the 
unicursal form that can be easily uniformized using a complex
variable $u$ as $v=w=u$. Then, the domain-wall solutions can be
expressed as rational or trigometric functions in $u$, and hence
$z$, when $u(z)$ is invertible in closed form. For genus 1, 
suitable bi-rational transformations to new variables $v(x,y)$, 
$w(x,y)$ will cast the curve into its standard Weierstrass form
\be
w^2 = 4v^3 - g_2 v - g_3 ~, 
\ee
which can be uniformized using the Weierstrass function
${\cal P}(u)$, with complex variable $u$ in the fundamental domain 
defined by the two periods $2 \omega_1$ and $2 \omega_2$ of the
surface, so that $v={\cal P}(u)$ and $w={\cal P}^{\prime}(u)$. 
Then, the corresponding domain-wall solutions can be described
explicitly in terms of elliptic functions, at least when 
$u(z)$ can be found in closed form. We will present 
some explicit examples of this type later. For models that 
correspond to higher genus surfaces, $g \geq 2$, the uniformization
is mathematically much more involved and will not be addressed here.

The application that we intend to consider in some detail in the
following concerns the spectrum of the quantum fluctuations for
the graviton as well as the scalar fields on the domain-wall 
backgrounds of gauged supergravity, which they turn out to coincide. 
Using the ansatz
\be
\Phi (x, z) = {\rm exp}(ik \cdot x) {\rm exp} \left(-{D-2 \over 2} A 
\right) \Psi (z)  
\ee
for a massless scalar field or any of the components of the graviton
tensor field, which represents plane waves propagating along the 
$(D-2)$-brane with an amplitude function that is $z$-dependent, 
we find that the spectrum of fluctuations is described by a 
one-dimensional
quantum mechanical problem. Setting $M^2 = -k \cdot k$ for the 
mass-square, we obtain a time-independent Schr\"odinger equation
for the function $\Psi (z)$, namely
\be
- \Psi^{\prime \prime} (z) + V(z) \Psi (z) = M^2 \Psi (z) ~. 
\ee
The potential is determined by the conformal factor of the metric
of the domain-wall background, $A(z)$, by
\be
V={(D-2)^2 \over 4} {A^{\prime}}^2 + {D-2 \over 2} A^{\prime \prime} ~,
\label{susy}
\ee
which is of the form appearing in problems of supersymmetric quantum
mechanics. 

The potential $V(z)$ can be determined explicitly once
the uniformization of the underlying algebraic curve has been carried
out in detail. The form can be rather complicated, depending on the
specific models, but for all of them the potential has the asymptotic form
\be
V(z) \simeq {D(D-2) \over 4} {1 \over z^2} ~, ~~~~ {\rm as} ~~ 
z \rightarrow 0^{+} ~, 
\ee
since the space approaches $AdS_D$ in the limit $z \rightarrow 0^{+}$ 
(or equivalently $F \rightarrow + \infty$). This means that the 
potential is unbounded from above. Its behaviour close to the other
end, namely $F \rightarrow b_{\rm max}$, depends on the multiplicity
of $b_{\rm max}$ in the algebraic form of the curve. A careful
analysis of the problem shows that if $b_{\rm max}$ appears $n$ times
(and so the model has an isometry group with an $SO(n)$ factor), the
potential will behave (including a subscript $n$ to distinguish among
various cases) as
\be
V_n \simeq {f_0^{1/2} \over 64} (3n^2 -8n)
(F-b_{\rm max})^{{1 \over 2}n -2} ~, ~~~~ {\rm as} ~~ 
F \rightarrow b_{\rm max} ~,  
\ee
where $f_0 = \prod_{i = n+1}^N (b_1 - b_i)$ (choosing 
$b_1 \equiv b_{\rm max}$) is a constant. Hence, for $n>4$, the potential
goes to zero at the other end and consequently the spectrum is
continuous. For $n=4$, the potential approaches a constant value, 
$f_0^{1/2}/4$,
and so the spectrum is again continuous but there is a mass gap whose
squared value is given by $f_0^{1/2}/4$. In both cases above, namely when
$n \geq 4$, the range of $z$ necessarily extends from 0 to $+ \infty$.
For $n<4$, on the other hand, the spectrum is discrete and $z$ extends
from 0 to some maximum (but finite) value $z_{\rm max}$, which is 
determined by the algebraic equation 
$F(z_{\rm max}) = b_{\rm max}$. In fact we find in this case that 
\be
V_n \simeq \left({(n-2)^2 \over (n-4)^2} - {1 \over 4} \right) 
{1 \over (z - z_{\rm max})^2} ~, ~~~~ {\rm as} ~~ z \rightarrow 
z_{\rm max}^{-}  
\ee
and the potential goes there to $+ \infty$ for $n=3$ and to $-\infty$ for
$n=1,2$. The latter two cases are not pathological since the coefficient 
of the $1/(z-z_{\rm max})^2$ term is larger or equal to $-1/4$ as required 
from elementary quantum mechanical considerations. Equivalently, 
since the potential
has the form \eqn{susy} appearing in supersymmetric 
quantum mechanics, the spectrum always has to be bounded from below
by zero. This completes the brief qualitative discussion of the
spectrum in all cases of interest.   
 
In supersymmetric quantum mechanics there is a superpotential $W(z)$ and
a pair of Schr\"odinger potentials associated to it
\be
V_- = W^2 - W^{\prime} ~,\qq V_+ = W^2 + W^{\prime} ~. 
\ee
Their spectrum are closely related to each other, and the same is true
for the corresponding eigen-functions, although there are some 
technical issues depending on whether supersymmetry is broken or not.
In our case the superpotential is provided by the conformal factor
of the metric, up to a scale, 
\be
W(z) = {D-2 \over 2} A^{\prime} (z) 
\ee
and the potential of the Schr\"odinger equation that describes the 
quantum fluctuations of the scalar and graviton fields on the 
domain-wall backgrounds of gauged supergravity (for $D=4$, 5 or 7)
is given by the form $V_+$. The partner potential $V_-$ sometimes
turns out to be easier to analyse quantum mechanically, 
although generically it does not itself 
correspond to the Schr\"odinger potential of a domain-wall solution
of gauged supergravity. Note at this point that if $A \rightarrow -A$
the superpotential will also flip sign and there will be an interchange
$V_+ \leftrightarrow V_-$. However, as $z \rightarrow 0$, 
the $AdS_D$ asymptotic behaviour of 
the domain-walls is not preserved under this interchange, and hence 
it can only be of mathematical interest for computing the spectrum
using $V_-$ instead of $V_+$. We will return to this later with 
specific examples. Finally, we mention that it is possible to apply
the WKB approximation, which works very well in supersymmetric quantum 
mechanics, in order to get results for the spectrum of quantum 
fluctuations on domain-wall backgrounds. Here, we will focus attention 
on the possibility to describe the spectrum exactly, in particular
for domain-walls associated to genus 1 Riemann surfaces where the
Schr\"odinger potential is expressed via the Weierstrass 
function in a generalized Lam\'e form. We will see that unlike other
problems in physics, where Lam\'e potentials with 
integer characteristics
become relevant, the characteristics turn out to be half-integer 
in gauged supergravity. Consequently, the spectrum is not of 
finite-zone type, and hence much more difficult to study exactly.
Thus, a number of questions will be left open for future study.

We proceed with the presentation of some explicit examples of 
elliptic type using the algebraic classification 
of domain-walls in terms of
Riemann surfaces and isometry groups.
In all these cases the Schr\"odinger potentials will assume the common
form
\be
V(z) = \mu(\mu+1) {\cal P}(z) + \nu(\nu+1) {\cal P}(z + \omega_1)
+ \kappa(\kappa +1) {\cal P}(z+ \omega_2) 
+ \lambda(\lambda +1) {\cal P}(z+ \omega_1 + \omega_2)  
\label{lame}
\ee
for different values of the coupling constants $\mu$, $\nu$, 
$\kappa$, $\lambda$ that will be determined in each case separately. 

\underline{$SO(2) \times SO(2) \times SO(2)$ in $D=5$}: The 
irreducible form of the algebraic curve is written directly in 
(hyper)-elliptic form
\be
y^2 = (x-b_1)(x-b_2)(x-b_3) 
\ee
and it can be brought into the standard Weierstrass form 
$w^2 = 4v^3 - g_2 v - g_3$ by the simple transformation
\be
y=4w ~, ~~~~~ x = 4v + {1 \over 3}(b_1 + b_2 + b_3) ~, 
\ee
in which case we find that
\ba
g_2 & = & {1 \over 36} \left((b_1 + b_2 - 2b_3)^2 - 
(b_2 + b_3 -2b_1)(b_1 + b_3 - 2b_2) \right) ~, \nonumber\\
g_3 & = & -{1 \over 432} (b_1 + b_2 - 2b_3) 
(b_2 + b_3 - 2b_1)(b_1 + b_3 - 2b_2) ~. 
\ea
Then, performing the uniformization in terms of the Weierstrass
function ${\cal P}(u)$ of a complex parameter $u$, we find
that $z=4 u$. The conformal factor of the metric turns out to be
\be
e^{2A(z)} = \left({1 \over 16} {\cal P}^{\prime} (u)\right)^{2/3} 
\ee
and so the corresponding domain-wall solution approaches $AdS_5$ (with radius
2) as $z \rightarrow 0^+$. On the other hand, since the
uniformizing parameter $u$ assumes real values from 0 to 
$\omega_1$ (real semi-period), we have that $z$ varies 
from 0 to $4 \om_1$.

In this case, the
corresponding potential in 
the Schr\"odinger differential equation for the variable $u$ is
\be
V(u) = {1 \over 4} \left(15 {\cal P}(u) - {\cal P}(u + \omega_1)
-{\cal P}(u + \omega_2) - {\cal P}(u + \omega_1 + \omega_2) 
\right)  
\ee
and so it has the generalized Lam\'e form 
\eqn{lame} with half-integer
coupling constants $\mu = 3/2$, $\nu = \kappa = \lambda = -1/2$. 
Using the identity
\be
4{\cal P}(2u) = {\cal P}(u) + {\cal P}(u + \omega_1) 
+{\cal P}(u + \omega_2) + {\cal P}(u+ \omega_1 + \omega_2) ~, 
\ee
it can be written into another form that we present here together with
the partner supersymmetric potential
\be
V_-(u) = 3{\cal P}(2u) ~, ~~~~~ V_+(u) = 4{\cal P}(u) -
{\cal P}(2u) ~. 
\ee
It is interesting to note in this case that the partner potential 
$V_-$ defines a simpler Schr\"odinger problem in the variable 
$\tilde{u} = 2u$ having $V(\tilde{u}) = n(n+1){\cal P}(\tilde{u})$ 
with $n=1/2$.

\underline{$SO(3) \times SO(3)$ in $D=5$}: In this case the 
algebraic curve of the model has the form
\be
y^4 = (x-b_1)^3 (x-b_2)^3 ~,  
\ee
which can be brought into the Weierstrass form
$w^2 = 4v^3 -g_2 v -g_3$ using the transformation
\be
x= b_1 + {1 \over v} \left(v + {1 \over 4}(b_2 - b_1)\right)^2 
~, ~~~~~ y = {w^3 \over 8v^3} ~, 
\ee
where 
\be
g_2 = {1 \over 4} (b_1 - b_2)^2 ~, ~~~~~ g_3 = 0 ~. 
\ee
Performing the uniformization in terms of the Weierstrass 
function ${\cal P}(u)$ of a complex variable $u$, we find
that $z = -8u$. Moreover, the conformal factor of the 
domain-wall metric is given by
\be
e^{2A(z)} = \left({{\cal P}^{\prime}(u) \over 
8 {\cal P}(u)}\right)^2 ~.  
\ee
Clearly, this conformal factor approaches
$4/z^2$ as $z \rightarrow 0$, which is the asymptotic
$AdS_5$ limit of the solution (with radius 2).    
In this case $z$ ranges from 0 to $8 \om_1$.
 
The Schr\"odinger potential for the corresponding differential equation written
using the variable $u$ turns out to be
\be
V(u) = {1 \over 4}\left(15 {\cal P}(u) + 3 {\cal P}(u +
\omega_1) + 3 {\cal P}(u + \omega_2) + 
15 {\cal P}(u + \omega_1 + \omega_2) \right) ~, 
\ee
which also has the generalized Lam\'e form \eqn{lame}  
with half-integer coupling constants
$\mu = \lambda = 3/2$ and $\nu = \kappa = 1/2$. 
Its supersymmetric partner can be easily determined
and turns out to be of the generalized Lam\'e form 
with $\mu = \lambda = 1/2$ and $\nu = \kappa = 3/2$.
These potentials exhibit a special invariance under
$u \rightarrow u + \omega_1 + \omega_2$ due to the
$SO(3) \times SO(3)$ symmetry of the underlying 
domain-wall solution.

\underline{$SO(2) \times SO(3)$ in $D=7$}: The genus 1 curve of 
this model is given by
\be
y^4 = (x-b_1)^2(x-b_2)^3 ~. 
\ee
It can be brought into the standard Weierstrass form
$w^2 = 4v^3 -g_2 v -g_3$ using the transformation
\be
x= {b_2 - b_1 \over 4} {w^2 \over v^3} + b_1 ~, ~~~~~ 
y= (b_2 - b_1) {w \over v} \left({w^2 \over 4v^3} 
-1\right) ~, 
\ee
where it turns out that
\be
g_2 = {1 \over 4} (b_1 - b_2) ~, ~~~~~ g_3 = 0 ~. 
\ee
We find that the uniformizing parameter $u$ is related to $z$ via
\be
u = \omega_1 + \omega_2 - {z \over 2} ~, 
\ee
whereas the conformal factor of the metric turns out to be
\be
e^{2A(z)} = \left({(b_1 - b_2)^2 \over 16} 
{{\cal P}^{\prime} (u) \over {\cal P}^3 (u)} \right)^{2/5} ~.  
\ee
The real variable $z$ ranges from 0 to $2 \omega_1$ if $b_1>b_2$ and 
from 0 to $2 (\om_1+\om_2)$ if $b_1<b_2$. The
domain-wall solution approaches asymptotically $AdS_7$ (with radius 4) as
$z \rightarrow 0^+$. 
 
The Schr\"odinger potential takes again the generalized 
Lam\'e form \eqn{lame}, after rescaling $z$ by a factor of 2, 
i.e. $z\to 2 z$,
\be
V(z) = {1 \over 4} \left(35 {\cal P}(z) - 
{\cal P}(z+ \omega_1) - {\cal P}(z + \omega_2) 
+ 3{\cal P}(z + \omega_1 + \omega_2) \right) 
\ee
with coupling constants $\mu = 5/2$, $\nu = \kappa = -1/2$, 
$\lambda = 1/2$. 
Its supersymmetric partner is
easily determined to be of the generalized Lam\'e type
with coupling constants $\mu = \lambda = 3/2$ and 
$\nu = \kappa = 1/2$, which concides with the 
Schr\"odinger potential for the quantum fluctuations
on the $SO(3) \times SO(3)$ domain-wall model in $D=5$
gauged supergravity.

\underline{$SO(2) \times SO(2) \times SO(2) \times SO(2)$ 
in $D=4$}: Another notable example is the genus 1 model
described by the algebraic curve in (hyper)-elliptic form 
\be
y^2 = (x-b_1)(x-b_2)(x-b_3)(x-b_4) ~. 
\ee
Setting $b_4 = 0$ without loss of generality, we employ 
the transformation
\be
x= {1 \over {1 \over 3}(b_1^{-1} + b_2^{-1} + b_3^{-1}) 
-v} ~, ~~~~~ y = {\sqrt{b_1 b_2 b_3} \over 2}
{w \over \left({1 \over 3}(b_1^{-1} + b_2^{-1} 
+b_3^{-1}) -v\right)^2} 
\ee
to bring the curve into its standard Weierstrass form 
$w^2 = 4v^3 -g_2 v - g_3$ with
\ba
g_2 & = & {2 \over 9}\left((b_1^{-1} + b_2^{-1} - 
2b_3^{-1})^2 + (b_2^{-1} + b_3^{-1} - 2b_1^{-1})^2
+ (b_3^{-1} + b_1^{-1} - 2b_2^{-1})^2 \right) ~, 
\nonumber\\
g_3 & = & {4 \over 27} (b_1^{-1} + b_2^{-1}
-2b_3^{-1})(b_2^{-1} + b_3^{-1} -2b_1^{-1})
(b_3^{-1} + b_1^{-1} - 2b_2^{-1}) ~.    
\ea
Then, uniformizing the curve, as usual, in terms of  
the Weierstrass function ${\cal P}(u)$ of a complex 
variable $u$, we find that $z$ is related to it by
\be
u = {\sqrt{b_1 b_2 b_3} \over 2} z + c ~; ~~~~ 
{\rm where} ~~ {\cal P}(c) = {1 \over 3}
(b_1^{-1} + b_2^{-1} + b_3^{-1}) ~. 
\ee
Also, the conformal factor of the corresponding 
domain-wall solution turns out to be
\be
e^{2A(z)} = {b_1 b_2 b_3 \over 4} 
\left({\cal P}(u - c) - {\cal P}(u + c) \right) ~, 
\ee
which approaches the $AdS_4$ limit $1 / z^2$ 
as $z \rightarrow 0^+$. It also turns out in this case
that $z$ ranges from 0 to a maximum value given 
by $2(\omega_1 -c)/\sqrt{b_1 b_2 b_3}$. 

The Schr\"odinger potential is calculated to be
\be
V(z) = {b_1 b_2 b_3 \over 4} \left(2{\cal P}(u+c)
+ 2{\cal P}(u-c) - {\cal P}(2u) \right) 
\ee
and it appears to be different from the generalized
Lam\'e form above. It is instructive, however, to
work out the supersymmetric partner potential. We
find in this case, when the 
Schr\"odinger equation is written in terms of the variable $u$, 
that the partners
are 
\be
V_- = 3{\cal P}(2u) ~, ~~~~~ 
V_+ = 2{\cal P}(u+c) + 2{\cal P}(u-c) - {\cal P}(2u) 
\ee
and so by an appropriate scaling of variable,  
$\tilde{u} = 2u$, the partner Schr\"odinger problem
has again the special Lam\'e potential
$V(\tilde{u}) = n(n+1){\cal P}(\tilde{u})$ with
$n=1/2$. The only difference from the previous 
case, where the $n=1/2$-Lam\'e potential makes its appearence, 
is that the  
variable $\tilde u$ ranges from $2c$ to $2\omega_1$, instead of 
the interval 0 to $2 \omega_1$.  
 
It will be interesting to revisit in future work 
the spectral properties
of the generalized Lam\'e potentials with half-integer 
coupling constants, in view of their relevance in theories
of gauged supergravity. There is only very little work on
this problem, which dates back to last century, and 
apparently turns out that 
such potentials are of infinite-zone type. Their 
structure becomes tractable when the underlying genus 1
Riemann surfaces degenerate by shrinking their $a$- or
$b$-cycles to zero size, in which case the exact spectrum is known
and coincides with the spectrum of the quantum Calogero
system. In this limit the potential becomes trigonometric and
hence the states are given in terms of elementary functions. 
For the elliptic models that arise here, however,
only the results from the WKB approximation are presently
known to the authors. 

Finally, another interesting problem
is the systematic construction of solutions in the
sector of gauged supergravity that also contains gauge fields.
The methods of algebraic geometry might prove again useful
for studying such generalizations.   

\vskip1cm
\centerline{\bf Acknowledgements}
\noindent
We are grateful to the organizers of the TMR meeting for their
kind invitation and generous financial support.

\end{document}